\documentclass[10pt,letterpaper]{article}
\usepackage[top=0.85in,left=2.75in,footskip=0.75in]{geometry}

\usepackage{amsmath,amssymb}

\usepackage{changepage}

\usepackage[utf8x]{inputenc}

\usepackage{textcomp,marvosym}

\usepackage{cite}

\usepackage{nameref,hyperref}

\usepackage[right]{lineno}

\usepackage{microtype}
\DisableLigatures[f]{encoding = *, family = * }

\usepackage[table]{xcolor}

\usepackage{array}

\newcolumntype{+}{!{\vrule width 2pt}}

\newlength\savedwidth

\raggedright
\usepackage[aboveskip=1pt,labelfont=bf,labelsep=period,justification=raggedright,singlelinecheck=off]{caption}

\makeatletter
\renewcommand{\@biblabel}[1]{\quad#1.}
\makeatother

\usepackage{lastpage,fancyhdr,graphicx}
\usepackage{epstopdf}
\fancyheadoffset[L]{2.25in}
\fancyfootoffset[L]{2.25in}
\def \tpT{{\mbox{\tiny {\sf T}}}}
\def \RR{{\mathbb{R}\,\!}}

\begin{document}
\vspace*{0.2in}

\begin{flushleft}
{\Large
\textbf\newline{
Trajectory planning optimization for real-time 6DOF robotic patient motion compensation
} 
}
\newline
\\
Rodney D. Wiersma\textsuperscript{1},
Xinmin Liu\textsuperscript{1}
\\
\bigskip
\textbf{1} Department of Radiation and Cellular Oncology, The University of Chicago, Chicago, IL 60637, USA
\\
\bigskip


\end{flushleft}
\section*{Abstract}
\textbf{Purpose:} Robotic stabilization of a therapeutic radiation beam with respect to a dynamically moving tumor target can be accomplished either by moving the radiation source, the patient, or both. As the treatment beam is on during this process, the primary goal is to minimize exposure of normal tissue to radiation as much as possible when moving the target back to the desired position. Due to the complex mechanical structure of 6 degree-of-freedom (6DoF) robots, it is not intuitive as to what 6 dimensional (6D) correction trajectory is optimal in balancing the shortest trajectory with the fastest trajectory. With proportional-integrative-derivative (PID) and other controls, potential exists that the controller may chose a trajectory that is highly curved, slow, or less optimal in that it leads to unnecessary exposure of healthy tissue to radiation. This work investigates a novel feedback planning method that takes into account a robot's mechanical joint structure, patient safety tolerances, and other system constraints, and performs real-time optimization to search the entire 6D trajectory space in each time cycle so it can respond with the most optimal 6D correction trajectory.

\textbf{Methods:} Computer simulations were created for two 6DoF robotic patient support systems: a Stewart-Gough plaform for moving a patient's head in frameless maskless stereotactic radiosurgery, and a linear accelerator treatment table for moving a patient in prostate cancer radiation therapy. The motion planning problem was formulated as an optimization problem and solved at real-time speeds using the L-BFGS algorithm. Three planning methods were investigated, moving the platform as fast as possible (platform-D), moving the target along a straight-line (target-S), and moving the target based on the fastest decent of position error (target-D). Both synthetic motion and prior recorded human motion were used as input data and output results were analyzed.

\textbf{Results:} For randomly generated 6D step-like and sinusoidal synthetic input motion, target-D planning demonstrated the smallest net trajectory error in all cases. On average, optimal planning was found to have a 45\% smaller target trajectory error than platform-D control, and a 44\% smaller target trajectory error than target-S planning. For patient head motion compensation, only target-D planning was able to maintain a $\leq$0.5mm and $\leq$0.5deg clinical tolerance objective for 100\% of the treatment time. For prostate motion, both target-S planning and target-D planning outperformed platform-D control.

\textbf{Conclusions:} A general 6D target trajectory optimization framework for robotic patient motion compensation systems was investigated. The method was found to be flexible as it allows control over various performance requirements such as mechanical limits, velocities, acceleration, or other system control objectives.

\section*{Author summary}

We present for the first time a general 6DoF trajectory planning method that can be used in real-time image guided radiation therapy procedures for robotic  stabilization of dynamically moving tumor targets. As the radiation beam is always on during the motion compensation process, it is mandatory that the 6D correction trajectory is optimal both spatially and temporally in order to maximize radiation to the tumor and minimize unintentional irradiation of healthy tissues. Unlike prior works, which relied on motion control approaches as PID or other controllers, this work presents the concept of motion planning, where all potential 6D trajectories are searched using ultrafast optimization methods and the best trajectory is chosen. As the method formulates the problem as an objective function to be solved, it allows high flexibility in that users can optimize various performance requirements such as mechanical robot limits, patient velocities, or other aspects that must operate within certain limits in order to ensure a safe medical process.


\section*{Introduction}

Modern radiation therapy (RT) delivery methods have evolved to the extent that technologies such as Intensity Modulated Radiation Therapy (IMRT) or Volumetric Modulated Arc Therapy (VMAT) can now tightly conform radiation to the 3D shape of a target with millimeter (mm) accuracy\cite{deodato2014stereotactic,otto2008volumetric}. This opens up the possibility of further radiation dose escalation to the tumor while still keeping doses low to surrounding organs-at-risk (OAR). However, as RT becomes increasingly conformal, and tends toward higher dose over fewer fractions, the issue of patient motion becomes ever more critical to address \cite{langen2001organ}. In the case of stereotactic radiosugery (SRS), or stereotactic body radiation therapy (SBRT), where small, highly focused, and accurate radiation beams are used to escalated doses in 1-5 treatments as opposed to smaller doses over 20-30 treatments, motion-related errors of even 1-2 mm can be significant.

Adaptive radiation therapy (ART) methods aim to address the RT motion management problem by using image guidance to locate the position of targets and then adapting the radiation dose to conform to these new positions. One form of ART is where dose re-planning is rapidly performed before the start of each treatment fraction while the patient is on the linear accelerator (LINAC). A daily cone-beam CT (CBCT) is taken by the LINAC, and deformable image registration (DIR) and dose recalculation is performed to correct the tumor and OAR volumes for shrinkage, expansion, motion, or other positional changes that may have occurred after the initial CT scan \cite{guckenberger2011potential}. A number of studies have evaluated the benefits of ART, indicating improved target coverage and reduced normal tissue toxicity
\cite{van2006conventional, kuo2006effect, yan2008developing, men2010gpu}. However, current ART modalities cannot address changes that may take place during actuation radiation beam delivery. Lung, prostate, pancreas, liver, and other thoracic and abdominal tumors have been shown to move as much as 35 mm with breathing, rectal filling, intestinal gas, or other types of biological motion \cite{davies1994ultrasound,ross1990analysis}. Clinical methods to manage such motion include beam gating, abdominal compress, or breath-hold \cite{kubo1996respiration, lohr1999noninvasive, kim2001held, wiersma2016high}. In recent years, significant research has been invested in exploring dynamical motion management strategies, such as moving the radiation source, the patient, or both \cite{d2005real, murphy2004tracking}. Here the primary goal is to stabilize the radiation beam with respect to a dynamically moving target. As the target can move along both translational (x, y, z) and rotational (pitch, row, yaw) axes as a function of time, precise positioning of the patient and LINAC becomes extremely challenging. This is especially true for single-isocenter-multiple-target SRS, where small rotations can lead to large positional errors for targets located away from isocenter, resulting in poor or missed target dose coverage and degraded treatment effectiveness \cite{redpath2008contribution, wang2008dosimetric, winey2014geometric, roper2015single, mancosu2015pitch, belcher2017patient}.

To effectively deal with intra-fractional patient motion, it is necessary to use a motion control algorithm that can response to both translational and rotational deviations in real-time. Various motion control algorithms have been investigated in RT, with the most focusing on real-time lung tumor motion compensation using a moving patient treatment table. D'Souza et al. employed a 1D controller that was used to drive a LINAC table such that it moves in phase with target, but in the opposite direction, to maintain the patient translational position \cite{d2005real}. Other methods used adaptive filter prediction and predetermined dynamic models to determine appropriate positioning commands \cite{wilbert2008tumor, menten2012comparison}. A proportional integral (PI) controller was developed and evaluated for a 1 degrees-of-freedom (DoF) treatment table tracking system to counter steer respiratory tumor motion \cite{lang2014development}. A coordinated dynamics-based proportional integral derivative (PID) control strategy was used to control the robotic system for continuously tracking translational position of the tumor \cite{buzurovic2011robotic}. The efficacy of this method was investigated by extensive computer simulation on two commercially available couches. Use of a linear Kalman filter to predict the surrogate motion with a linearized state space model to predict table position and velocity were investigated \cite{haas2012couch}. A model predictive control (MPC) of a robotic treatment table for motion compensation was reported in \cite{herrmann2011model}. To maintain dynamic behavior of systems in the face of perturbations and other uncertainties, a robust control for parallel robotic platforms was investigated and compared with a widely used PID approach using extensive computer simulations \cite{buzurovic2012robust}. For robotic SRS based head motion with angular stabilization, a decoupling control method for a 4D (xyz+pitch) robot was investigated \cite{liu2015robotic, liu2015roboticACC}.

As the radiation beam is on during real-time patient 6DoF motion compensation, it is mandatory that the target correction trajectory is optimal both spatially and temporally in order to reduce unnecessary exposure of healthy tissue to radiation. Most prior investigations were limited to only 3DoF translational (xyz) motion, and therefore are not suitable to 6DoF robotic systems \cite{belcher2017towards}. Applying such methods to a 6DoF robot can result in a correction that may reach the desired position following a path that is slow, highly curved, or less optimal. Additionally, most prior works have focused on respiratory motion compensation, and have employed the use of prediction algorithms to buffer against system lag times. Such methods are therefore not suitable for prostate, liver, head, or other types of motion that can be highly unpredictable \cite{wiersma2009development,liu2018general}. To address these issues, we have developed for the first time, a novel optimization based motion controller that takes into account both robot mechanical constraints and patient safety tolerances, and performs a search over all possible 6D correction trajectories in order to select one that is most optimal. The method is universal as it can be applied to any robotic system provided that the robot's joint configuration is well-known. Since the method formulates the correction trajectory problem as an objective function to be optimized, various constraints can be easily applied on actuator mechanical limits, patient velocities, and other aspects of the system that must operate within fixed limits during the motion compensation process. We demonstrate that standard convex optimization methods can be used to solve this problem at real-time speeds. As the controller responds with the most optimal path, robot lag time is also reduced, allowing for motion compensation in cases of slow unpredictable motion without the use of prediction. In cases of rapid, but predictable motion, such as lung based tumors or other respiratory coupled targets, the lower lag time allow less prediction to be used. This is beneficial from a patient safety standpoint, where prediction accuracy deteriorates the further a prediction algorithm must predict in the future.

\section*{Materials and methods}

\subsection*{Robotic system simulations}

\begin{figure*}[ht]
	{\bf (a)  \hspace{60mm} (b) \hspace{25mm} } \newline
	\centerline{
		\includegraphics[height=45mm]{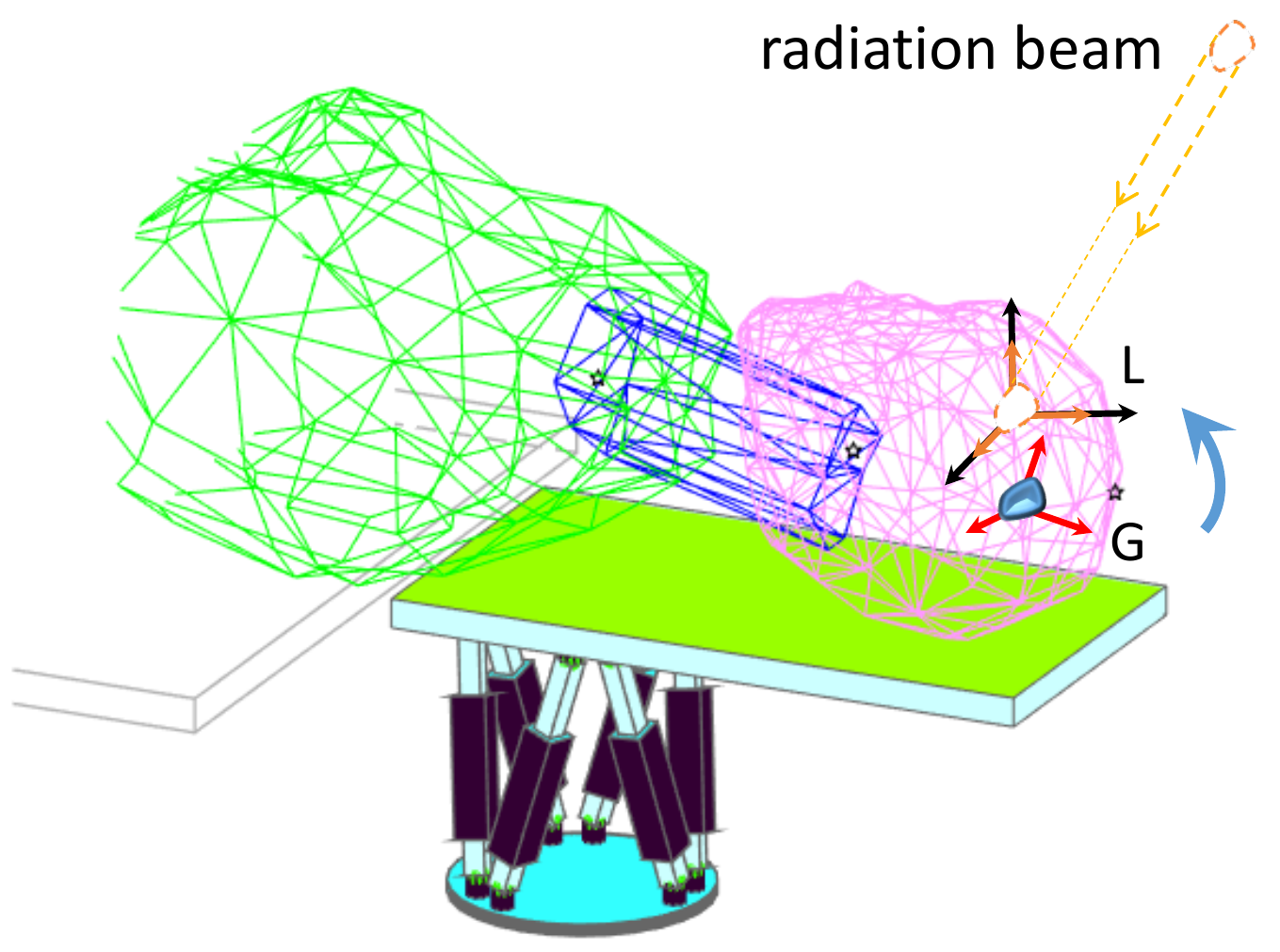} \hspace{7mm}
		\includegraphics[height=45mm]{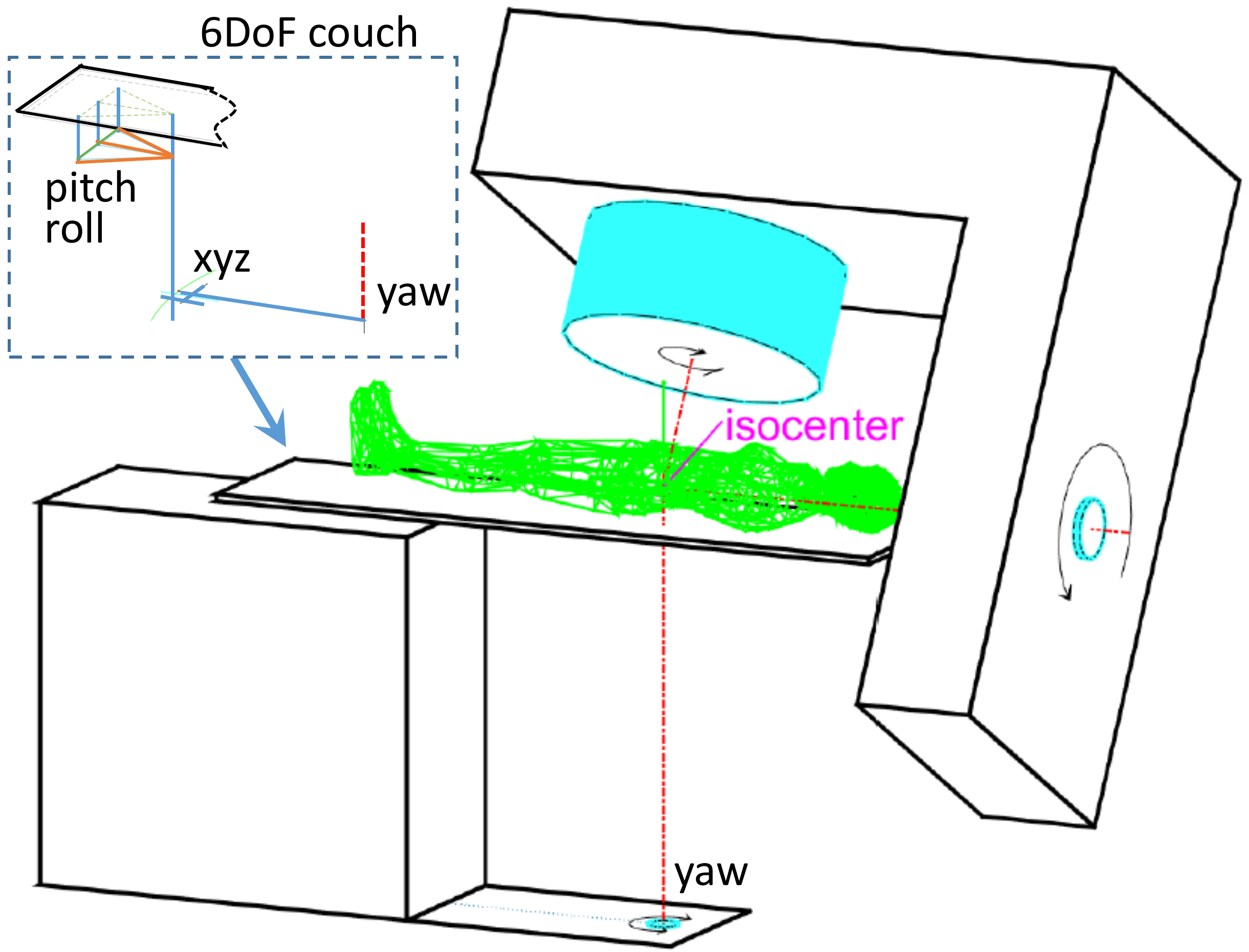}
	}
	\vspace{2mm}\newline
	\centerline{ \bf (c) \hspace{120mm} }\newline \vspace{-8mm}\newline
	\centerline{\includegraphics[height=49mm]{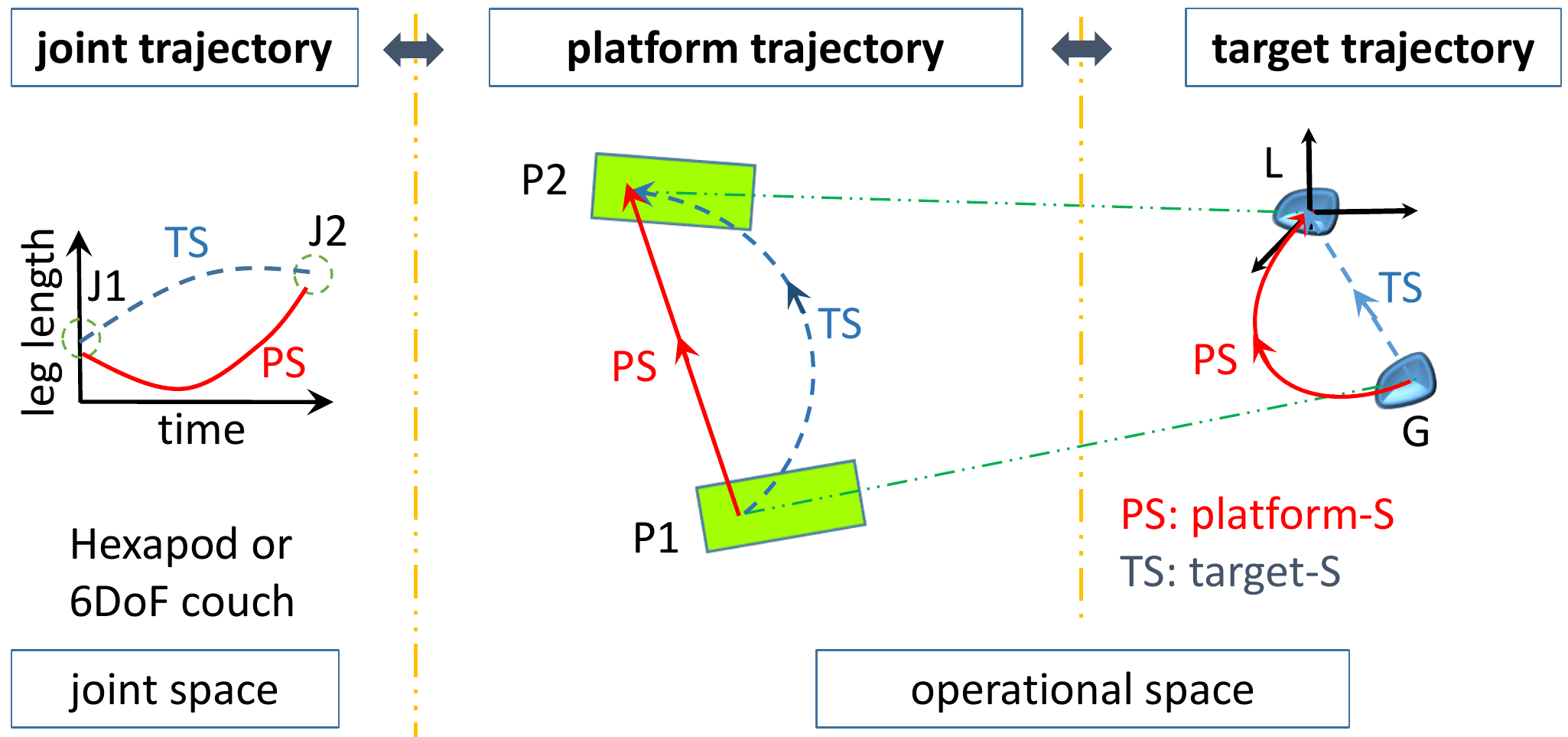} }
	\caption{
		(a)(b) Simulations of two 6DoF motion compensation systems for radiation therapy. (c) Three types of trajectories involved in the motion compensation: joint, platform, and target. To send the target from $G$ back to the desired position $L$ in operational space, a motion of the platform from $P1$ to $P2$ is needed, and such a motion can be implemented by changing the actuator length from $J1$ to $J2$ in the joint space. As shown, performing a straight-line trajectory in one space can result in a highly curved trajectories in other spaces.
	}
	\label{fig_hexapod_couch}
\end{figure*}

Similar to other works that have investigated robotic motion control, we employ the use of computer simulation to test various kinematic and dynamic properties of the robotic system \cite{buzurovic2011robotic, haas2012couch}. As shown in Figure~\ref{fig_hexapod_couch}, dynamic computer simulations of a 6DoF SRS system and a 6DoF LINAC treatment table were constructed using Matlab (Mathworks, Natick, MA). The SRS simulation consists of six linear actuators arranged in a parallel kinematic fashion that performs real-time head motion stabilization by moving a platform (end-effector) that supports the patient's head \cite{belcher2016spatial,belcher2014development}. The treatment table simulation is modeled after a commercial product (PerfectPitch, Varian Medical System, CA), and is a serial-parallel kinematic system where the platform moves the entire patient.

The clinical target to be stabilized with respect to the treatment beam will be considered as a rigid body and thereby represented by 6D coordinates that include the linear position (x, y, z) and the orientation (pitch, roll, yaw). The target is located at position $G$ in the LINAC coordinate frame, and is tracked in real-time during the treatment using a suitable patient motion monitoring device such as kV fluoroscopy, infrared (IR) markers, or 3D surface imaging \cite{wiersma2009development, wiersma2013spatial, grelewicz2014combined}. The desired position (setpoint) is given as $L$, and corresponds to LINAC isocenter. As shown in Figure ~\ref{fig_hexapod_couch}, the motion control problem can be divided into two spaces: joint (robot actuators) and operational (platform and target). In the simplest case, where a robot's axes are all aligned with the LINAC frame, and one only considers 3D translational motion (xyz), the joint, platform, and target will all move along the same trajectory. In this case one does not require trajectory planning, as a straight-line trajectory from $G$ to $L$ will always be optimal. However, with a 6DoF robot, that is capable of both translational and rotational motion, the joint, operational, and target trajectories may differ substantially. For example, a straight-line target trajectory (blue dashed line), may result in highly curved joint and platform trajectories. Such a trajectory may take significant time to complete as all actuators must now travel large distances. It is no longer intuitive as to what 6D trajectory is now optimal both spatially and temporally.

\subsection*{Robotic motion compensation scheme}

\begin{figure}[ht!]
	\centerline{\includegraphics[width=60mm]{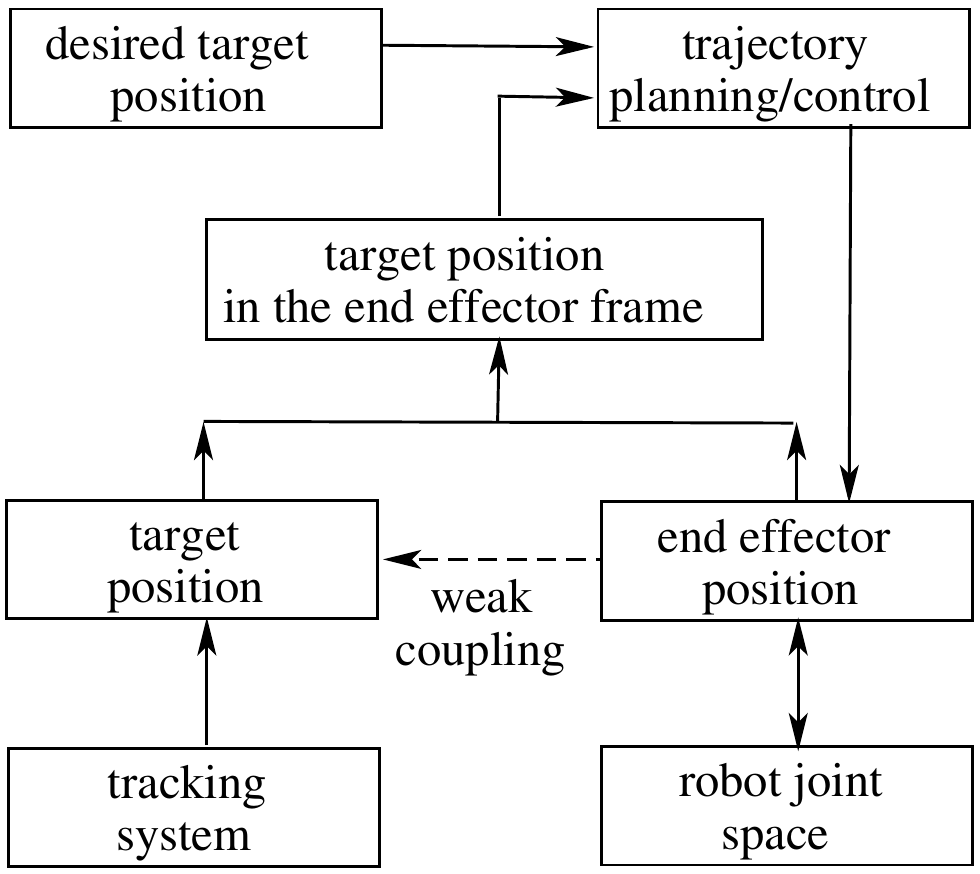}}
	\caption{ Signal flow diagram of the robotic motion compensation system.}
	\label{fig:signal_flow}
\end{figure}

For a given measured target position with a displacement away from the desired setpoint, the first step is to compute what is the required robot platform position for returning the target back to the setpoint. This can be done by the signal flow diagram in Figure~\ref{fig:signal_flow}. Suppose the measured target position is $(r_g, \psi_g)$, the position of the platform is $(r_u, \psi_u)$, and the position of the target in the platform frame is $(r_{gu}, \psi_{gu})$. Then,

\begin{equation}\label{6D_g}
r_g=r_u+\Omega(\psi_u)^\tpT r_{gu}, \qquad
\Omega(\psi_g)=\Omega(\psi_{gu}) \Omega(\psi_u),
\end{equation}
where $\Omega(\cdot)$ is the direction cosine matrix, and ($r_{gu}, \psi_{gu}$) is the target position in the platform coordinate frame. The position ($r_{gu}, \psi_{gu}$) can be calculated by using the current platform position and the measured target position,
\begin{equation}\label{6D_32}
r_{gu}=\Omega(\psi_u) (r_g-r_u), \qquad
\Omega(\psi_{gu})= \Omega(\psi_g)\Omega(\psi_u)^\tpT.
\end{equation}
Assume that the target position with respect to the platform remains unchanged during each robot time cycle, it can be verified that by moving the platform to the following position $(\hat r_u, \hat \psi_u)$,
\begin{equation}\label{6D_u}
\hat r_u=\hat r_g-\Omega(\hat \psi_u)^\tpT r_{gu}, \qquad
\Omega(\hat \psi_u) =\Omega(\psi_{gu})^\tpT \Omega(\hat \psi_g),
\end{equation}
the target will move back to the desired position.

The actuator length of the robot can be calculated accordingly based on the platform position $(r_u, \psi_u)$ by inverse kinematics. Suppose the coordinates of the robot joints in the LINAC coordinates frame is $D_{wi}\in\RR^3$, and the coordinates of the corresponding joints at in the platform frame is $D_{ui}\in\RR^3$. Then the actuator length is given by

\begin{equation}\label{lt}
\ell_i=\|r_u+\Omega(\psi_u)^\tpT D_{ui}-D_{wi}\|,
\end{equation}
for $i=1,2,\cdots, 6$.

For easy reference, let us make the following denotations. Denote the 6D target and platform position vectors as $x:=(r_g; \psi_g)$ and  $u:=(r_u; \psi_u)$, respectively. Here $(y;z)$ is used to denote the column vector $(y^\tpT \; z^\tpT)^\tpT$. Denote initial positions as $x_o=x(0)$ and $u_o=u(0)$, initial actuator length as 6D vector $\ell_o=\ell(0)$. Denote the desired target position as $\hat x:=(\hat r_g; \hat \psi_g)$, the required position of platform that pushes the target back to the desired position as $\hat u:=(\hat r_u; \hat\psi_u)$, and the corresponding actuator length as $\hat \ell$.

\subsection*{Optimization based trajectory planning}

Once the start and desired positions of the platform are known, the next step is to design a suitable trajectory between these two points. For reference, a path denotes the locus of points in a space and is a pure geometric description of motion, while a trajectory is a path with timing information \cite{siciliano2010robotics}. For robotic motion compensation it is critical that the target trajectory converges to the desired position in an optimal way, such that a minimum time, shortest path, or steepest descent of positional error is achieved. Failure to do so will lead to unnecessary exposure of healthy tissue to radiation and poor tumor dose conformality.

To solve this problem, trajectory planning will be considered by three trajectories in two spaces: joint trajectory in the joint space, and platform trajectory and target trajectory in the operational space (Figure~\ref{fig_hexapod_couch}). When the platform moves along a trajectory in the operational space, $u(k), k=0,1,\cdots,n$ with $u(0)=u_o$ and $u(n)=\hat u$ for certain $n$, the target will approach the desired position following a certain trajectory in the operational space, $x(k), k=0,1,\cdots,n$ with $x(0)=x_o$ and $x(n)=\hat x$. To implement such an platform trajectory, the actuator length should thus follow a certain trajectory in the joint space, $\ell(k), k=0,1,\cdots,n$ with $\ell(0)=\ell_o$ and $\ell(n)=\hat \ell$.

Due to the many parameters involved, it is not intuitive as to what trajectory is optimal both spatially and temporally. As can be seen in Figure~\ref{fig_hexapod_couch}, following the shortest platform path (platform-S) is not necessarily the best path for the target. Fortunately, such problems can be solved efficiently in real-time by use of optimization algorithms such as the quasi-Newtonian Limited-Broyden-Fletcher-Goldfarb-Shannon (L-BFGS) algorithm \cite{morales2011remark}, Proximal Operator Graph Solver (POGS) \cite{parikh2014block,fougner2015parameter,liu2016constrained,liu2017use}, or other fast optimization algorithms. As a rule, the correction trajectory must meet control objectives and are subject to the robot's mechanical constraints and system's dynamic constraints. The robot's actuator constraints can be specified in joint space as,
\begin{equation}\label{ell1}
-\Delta\ell_{\max} \preceq \ell(k+1)-\ell(k)\preceq \Delta\ell_{\max},
\end{equation}
where $\Delta\ell_{\max}$ is the limit of actuator length change in each step.
And the dynamic constraints can be specified in operational space as
\begin{equation}\label{vmax}
-v_{\max} \preceq u(k+1)-u(k) \preceq v_{\max},
\end{equation}
where the vector $v_{\max}$ is the maximum 6DoF motion of the platform in one step.

By use of (\ref{lt}), the constraint (\ref{ell1}) can be represent as the following platform motion constraints,
\begin{equation}\label{ell2}
-\Delta\ell_{\max} \preceq  B (u(k+1)-u(k)) \preceq \Delta\ell_{\max},
\end{equation}
where $B$ is the Jacobian matrix that can be considered as a constant matrix in each time cycle and is computed based on $x(k)$ and $\ell(k)$.

To facilitate optimal target trajectory planning, the robot system is first discretized. Assume that the target position keeps unchanged with respect to the platform, {\em i.e.}, $(r_{gu}, \psi_{gu})$ is constant, then (\ref{6D_g}) can be represent as
$
x=f(u),
$
where $x$ and $u$ are the 6DoF position vectors of the target and the platform, respectively.
The discrete system is given by
\begin{equation}\label{nonlinear_syms}
x(k)=f(u(k)).
\end{equation}

A general optimal target trajectory planning can be formulated as
\begin{equation}\label{general_opt}
\begin{array}{rl}
	\mbox{minimize} &\sum_{k=0}^\infty h(x(k),\hat x, u(k)),\\
	\mbox{subject to} & (\ref{vmax}), \; (\ref{ell2})\; \mbox{and} \; (\ref{nonlinear_syms}).
\end{array}
\end{equation}
The nonlinear optimization problem can be approximated as a problem on linear systems. The linearization is performed in each time cycle, so it can describe the system in each small motion accurately. Consider $x(k+1)=f(u(k+1))$ and small platform motion in each step $v(k)=u(k+1)-u(k)$.
By linearization, the motion compensation discrete system is given by
\begin{equation}\label{xk1}
x(k+1) = x(k) +A v(k),
\end{equation}
where the matrix $A$
\begin{equation}\label{mat_A}
A=\frac{\partial f} {\partial u}\Bigl|_{r_u, \psi_u, r_{gu}, \psi_{gu}(k)}
\end{equation}
depends on both $(r_u, \psi_u)$ and $(r_{gu}, \psi_{gu})$ of moment $k$, and can be updated in each step.

The target can move back to the desired position following a straight-line in the target space, and such a trajectory can be obtained by defining an appropriate $h$ in (\ref{general_opt}).
This path is interpolated between $x_o$ and $\hat x$,
\begin{equation}\label{mu}
x(k)=x_o+(\hat x-x_o) \mu,
\end{equation}
where the scalar $0\leq \mu \leq 1$ is $k$ dependent, and $\mu$ can be maximized for a quick motion compensation to the desired target position $\hat x$. The optimization can be implemented step-by-step. For simplification, consider only mechanical constraints (\ref{ell2}). In each step, let $x_o=x(k-1)$. The desired next target position is (\ref{mu}), and the required platform motion is $v=A^{-1}(\hat x- x_o)\mu$. The scalar $\mu$ should be found to minimize
$
h=(x(k)-\hat x)^\tpT  (x(k)-\hat x)= (\mu-1)^2 (x_o-\hat x)^\tpT (x_o-\hat x)
$. This is equivalent to minimize $(\mu-1)^2$,
subject to $ - v_{\max}  \preceq A^{-1}(\hat x- x_o)\mu \preceq  v_{\max}$.
Thus, it can be verified that the fastest straight target trajectory is given by
\begin{equation}\label{opt_TS}
\begin{array}{ll}
	x(k+1)=x(k)+(\hat x-x(k)) \mu, \\
	\hspace{5mm} \mu=\min(1,1/\max |(A^{-1}(\hat x-x(k)))/v_{\max}|).
\end{array}
\end{equation}
Refer to such target straight-line planning as {\em target-S}. In target-S, the target follows a straight line for the start point to the end point, and the required platform and actuator length follow curve lines in operational space and joint space, respectively (Figure~\ref{fig_hexapod_couch}).

As forcing the robot to move the target along a straight-line path can lead to over constraining the system, it does not fully exploit the robot's potential in reducing the target's position error in a temporally optimal way. To include both spatial and temporal components in the optimization, a steepest descent of target position error in each step can be used by formulating (\ref{general_opt}) as follows,

\begin{equation}\label{opt_TD}
\begin{array}{ll}
	\mbox{minimize} & (x_o+Av-\hat x)^\tpT (x_o+Av-\hat x),\\
	\mbox{subject to} & - v_{\max}  \preceq v \preceq  v_{\max}. 
\end{array}
\end{equation}

For each step, $x_o$ is updated, $x_o=x(k-1)$, and according to (\ref{xk1}), the next step target position is $x(k)=x_o+Av$, where the vector $v=u(k)-u(k-1)$ is the platform motion to be optimized. Refer the planning (\ref{opt_TD}) as {\em target-D planning}.

\subsection*{Evaluation of robotic compensation for standard motions}

The robotic SRS system was evaluated by two synthetic motion standards: step-like motion and oscillating motion. Step motion was simulated since it can represent system response performance to a sudden target change. Sinusoidal motion was considered since it evaluates how well the system responses to dynamical motion and is closely related to target oscillation caused by respiratory motion.

\subsection*{Evaluation of robotic compensation for volunteer head motion data}

Both the robotic SRS system and 6DoF treatment table were tested using prior recorded patient data. Real-time 6D head positional data was obtained from six volunteers under a IRB approved study (IRB14-0535) at the University of Chicago \cite{belcher2017patient}. All volunteers were asked to rest in a comfortable supine position on a head support without the use of a mask or any additional immobilization. A virtual target was selected in the parietal lobe, and was tracked through IR makers fixed to a frame rigidly attached to the head. For each experiment, target motion data were recorded over 15 minutes at 12 frames per second. The recorded target motion data was used as input motion for the optimization based trajectory planning algorithm. Real-time 6D prostate motion was obtained was obtained from Tehrani et al \cite{tehrani2013real}.

\section*{Results}

\subsection*{Step-like and sinusoidal target motion}

\begin{figure}[h!]
	{\hspace{2mm} \bf (a)  \hspace{80mm} (b) \hspace{70mm} $\;$}\newline
	\centerline{
		\includegraphics[width=80mm]{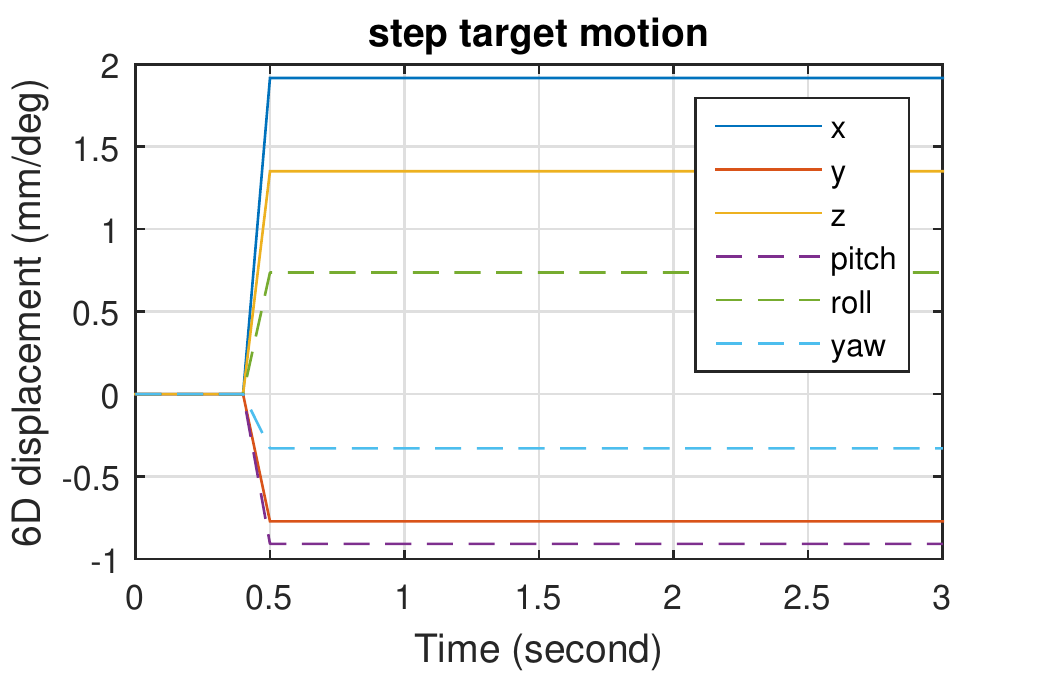}
		\includegraphics[width=80mm]{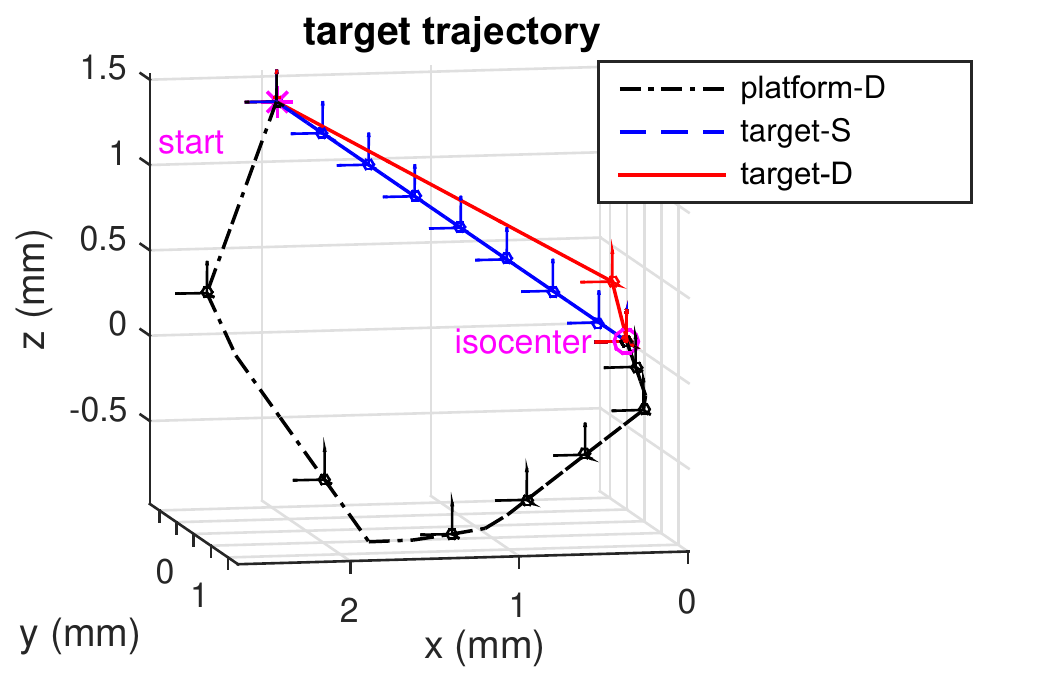}
	}\newline \vspace{10mm}\newline
	{\hspace{30mm} \bf (c)  \hspace{80mm} (d) \hspace{70mm} $\;$}\newline
	\centerline{
		\includegraphics[width=80mm]{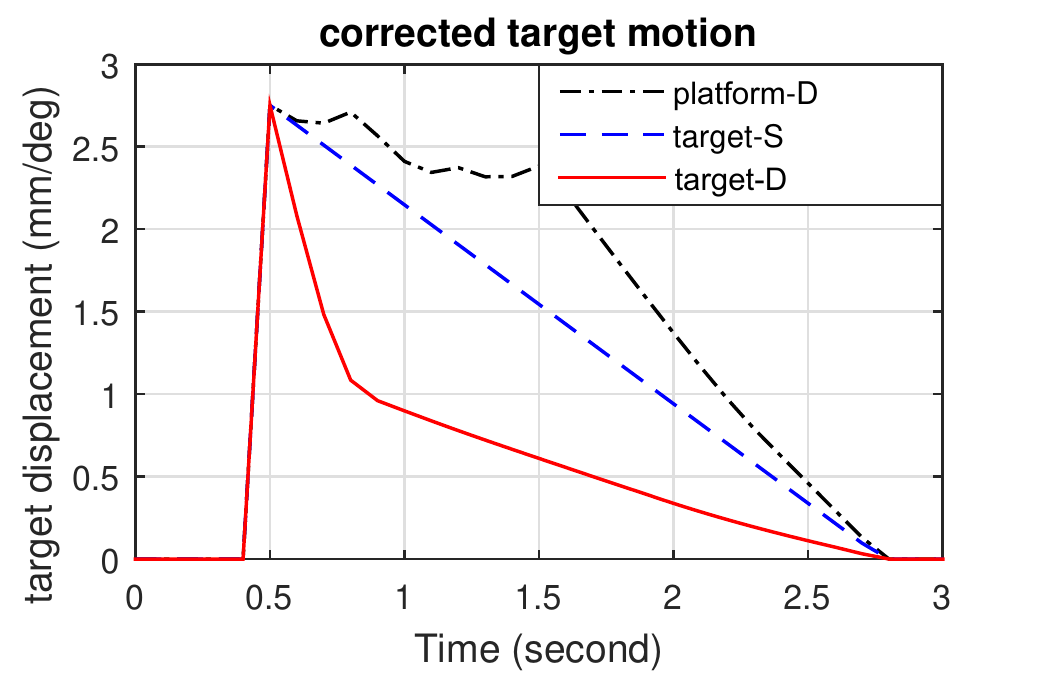}
		\includegraphics[width=80mm]{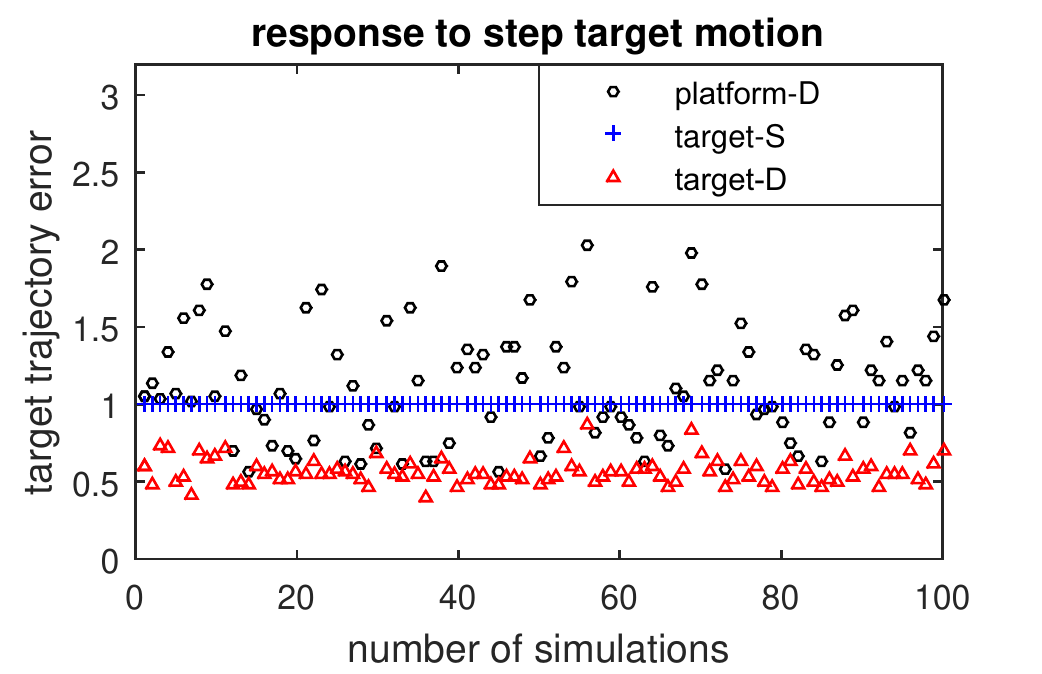}
	}
	\caption{
		(a)(b)(c): System response to step like target deviations. (a) Input motion. (b) Target trajectory with orientation shown every 0.15 seconds. (c) Target displacement versus time. (d) Integrated trajectory errors for 100 simulations using randomly generated 6D input motion within 2mm/1deg. Errors were normalized to target-S planning case.
	}
	\label{fig_sim_DSD1}
\end{figure}

The simulated robotic SRS system was tested against sudden step-like displacements of the target away from the setpoint (Figure~\ref{fig_sim_DSD1} (a)). System response to this motion using platform-D control, target-S planning and target-D planning is shown in Figure~\ref{fig_sim_DSD1} (b). Here each point represents a 0.15 s time interval and the vector arrows attached to each point denote rotations. Although all planning strategies eventually converged to the setpoint, it can be seen that certain strategies were more appropriate to motion compensation in RT. For platform-D planning (black line), the target can deviate significantly from a straight-line path and can take a long time to converge. In this case, the target temporally moved further away and took approximately 1.6 s to reach a value less than 1 mm/deg away from the setpoint. From a target perspective, such a trajectory is not optimal as it would lead to unnecessary irradiation of healthy tissues. For target-S planning (blue line), the target moved at a constant speed and followed the expected straight-line path. However, although optimal spatially, this trajectory is not optimal temporally as it takes approximately 1.5 s to reach less than 1 mm/deg away from the setpoint. For target-D planning (red line), the target followed a path that was close to a straight-line and converged to the setpoint within 0.3 s. In this case, both spatial and temporal components of the trajectory were optimal. The time course $\|x(k)-\hat x\|$ of the three trajectory planning strategies is given in Figure~\ref{fig_sim_DSD1} (c).

To test system response against many arbitrary 6D directions, 100 randomly generated target step-like displacements with amplitudes within a 2mm/1deg were inputted into the simulated robotic SRS system. The performance of different control/planning methods was evaluated by the target trajectory error defined as
\begin{equation}\label{planning_error}
E=\sum_{k=1}^\infty \|x(k)-\hat x\|,
\end{equation}
where the difference from the desired setpoint is summed over all time steps. The target trajectory errors of 100 simulations are  plotted in Figure~\ref{fig_sim_DSD1} (d). The trajectory errors of target-S planning were smaller than those of platform-D control in 56 cases, and target-D planning had the smallest trajectory error in all 100 cases. On average, target-D planning was found to have a 45\% smaller target trajectory error than platform-D control, and a 44\% smaller target trajectory error than target-S planning.

\begin{figure*}[ht!]
	{\hspace{2mm} \bf (a)  \hspace{80mm} (b) \hspace{70mm} $\;$}\newline
	\centerline{
		\includegraphics[width=80mm]{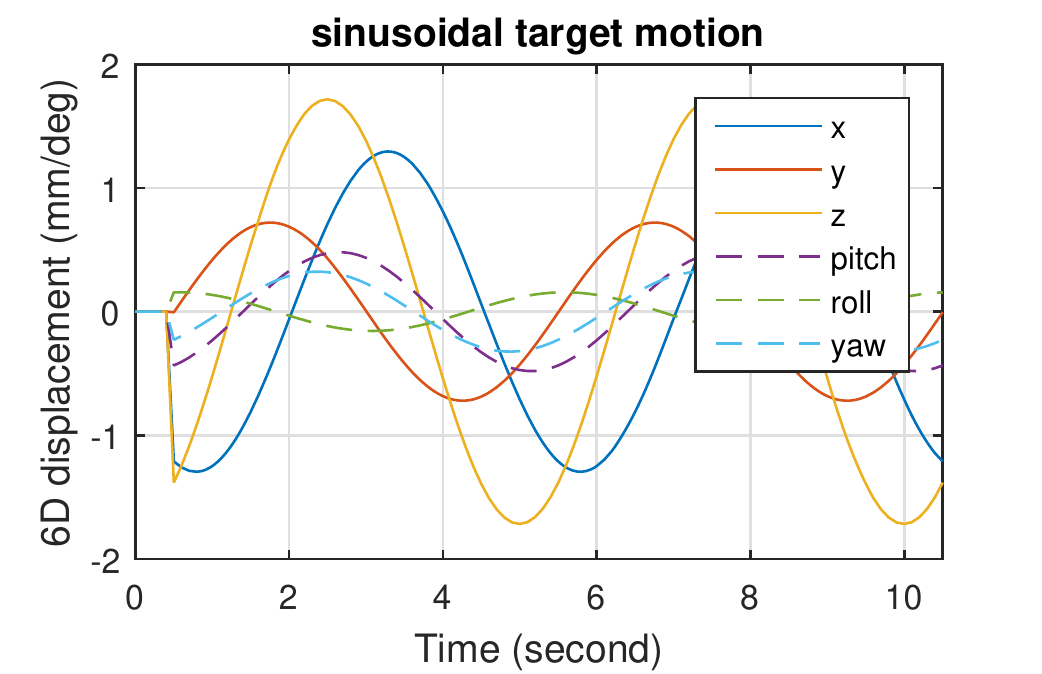}
		\includegraphics[width=80mm]{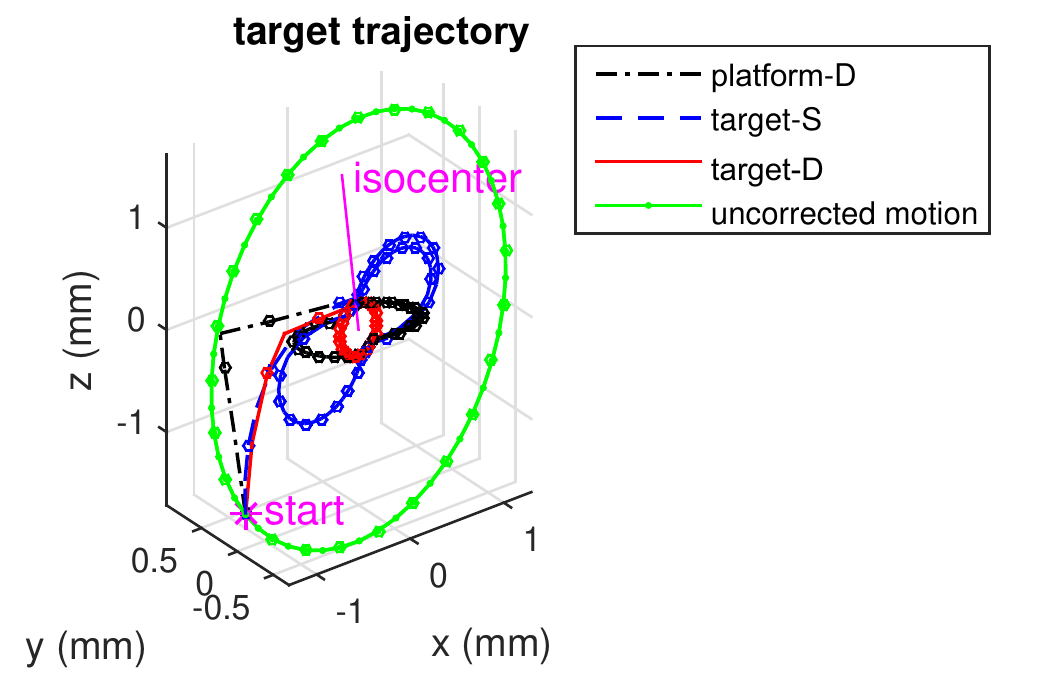}
	}\newline
	\vspace{20mm}\newline
	{\hspace{30mm} \bf (c)  \hspace{80mm} (d) \hspace{70mm} $\;$}\newline
	\centerline{
		\includegraphics[width=80mm]{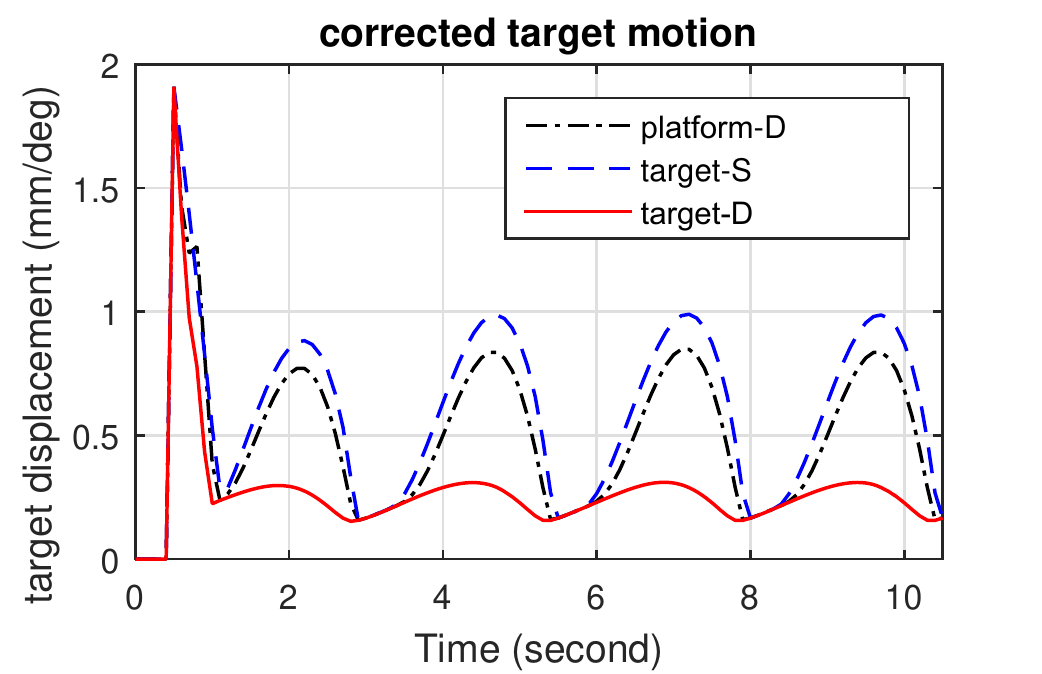}
		\includegraphics[width=80mm]{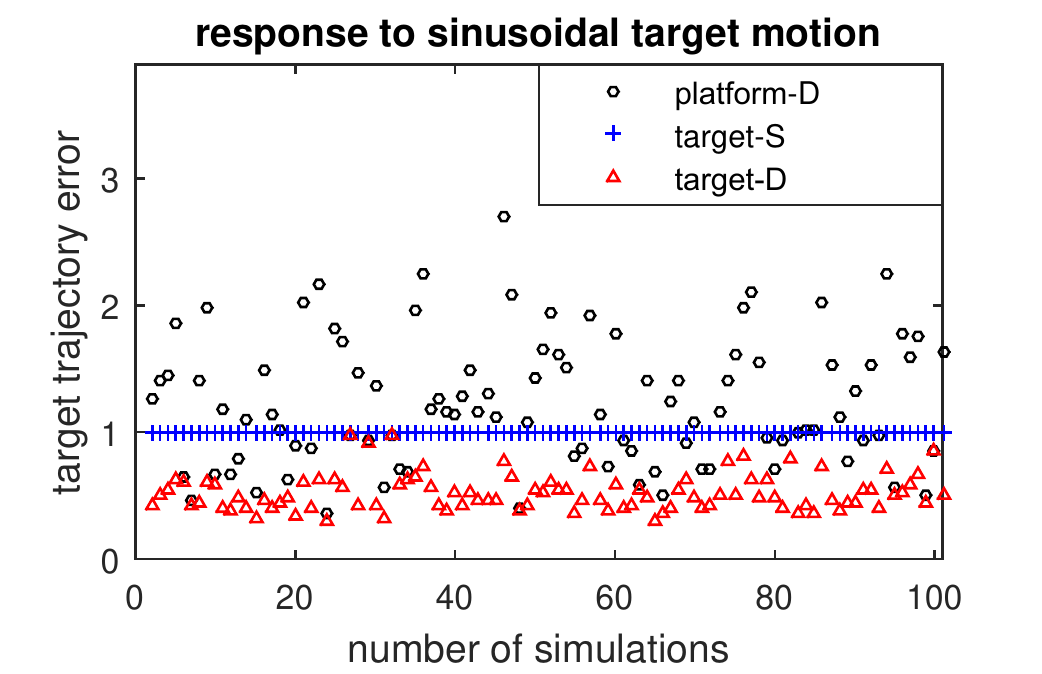}
	}
	\caption{
		(a)(b)(c): System response to sinusoidal target motion deviation. (a) Input sinusoidal target motion. (b) Target trajectory shown every 0.2 seconds. (c) Target displacement versus time. (d)  Integrated target trajectory errors for 100 simulations using randomly generated input motion within 2 mm / 1 deg. Errors were normalized to target-S planning case.
	}
	\label{fig_sim_sin1}
\end{figure*}

The performance of optimization based trajectory planning was also evaluated for sinusoidal target motion around the desired setpoint. Fundamentally, due to lag time between measurement and robot actuation, the corrected target position will not remain at the setpoint, but rather, the target will move around the setpoint with a smaller amplitude of oscillation. This amplitude can therefore be used as an indicator of the efficiency of the motion correction algorithm. The system response of platform-D control, target-S planning and target-D planning for a randomly generated 6D target oscillation is shown in Figure~\ref{fig_sim_sin1} (a)(b)(c). The trajectory of target-D planning was found to converge to a steady oscillation faster than platform-D control and target-S planning, and also showed a smaller oscillation amplitude. Figure~\ref{fig_sim_sin1} (d) shows the resultant errors for 100 simulations using randomly generated sinusoidal input motion with amplitudes within 2mm/1deg. Target-S planning has smaller target trajectory errors than platform-D control in 68 cases, and target-D planning has smallest target trajectory errors in all 100 cases. On average, target-D planning has 52\% smaller target trajectory error than platform-D control, and has 48\% smaller trajectory error than target-S planning.

\subsection*{Human volunteer motion}

\begin{figure*}[ht!]
	\centerline{\sf\footnotesize \underline{\hspace{5mm} uncorrected target motion \hspace{5mm}}}
	\centerline{
		\includegraphics[width=70mm]{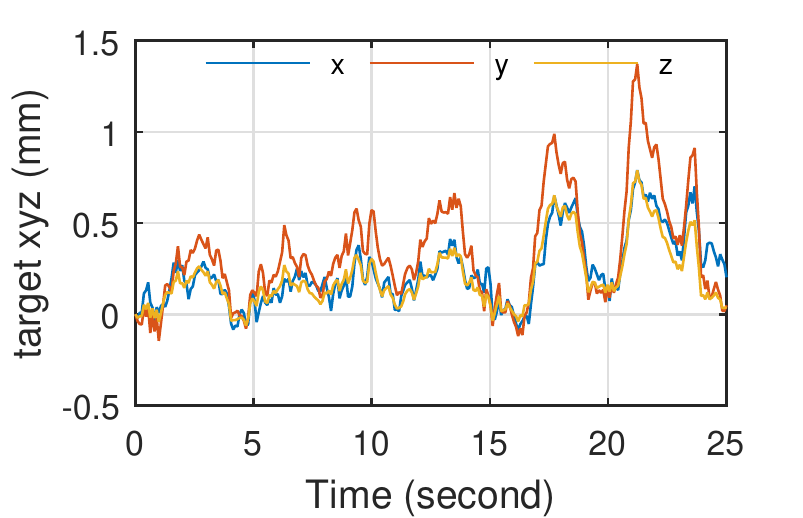}
		\includegraphics[width=70mm]{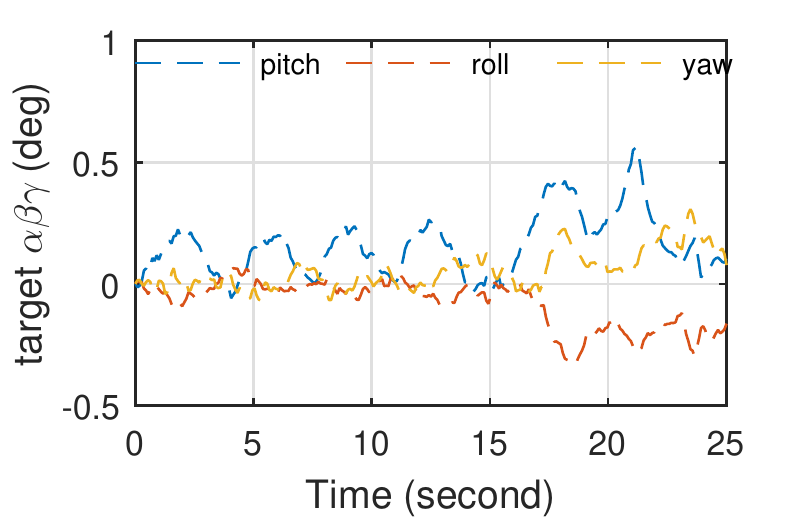}
	}
	\centerline{\sf\footnotesize \underline{\hspace{15mm} corrected target motion by different planning (mm/deg)\hspace{15mm}}}
	\centerline{
		\includegraphics[width=57mm]{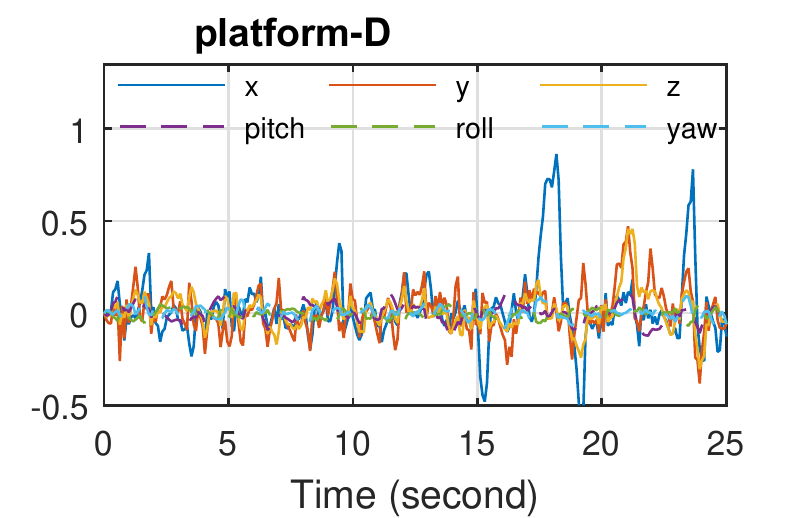}\hspace{-5mm}
		\includegraphics[width=57mm]{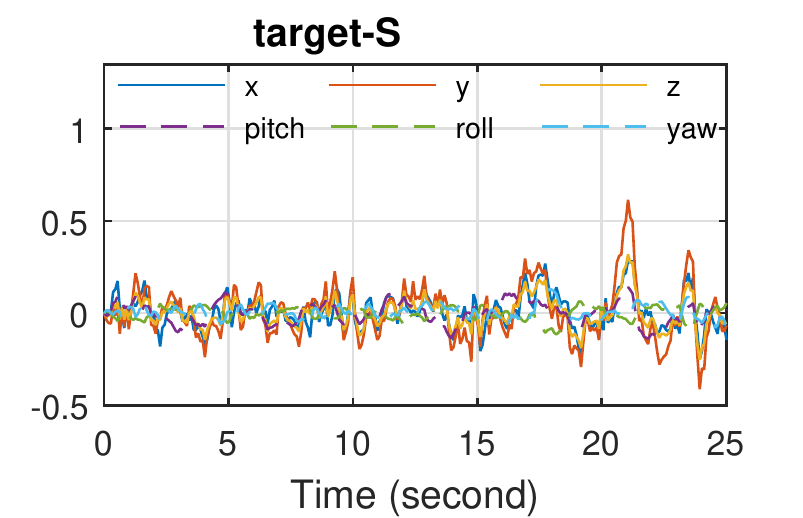}\hspace{-5mm}
		\includegraphics[width=57mm]{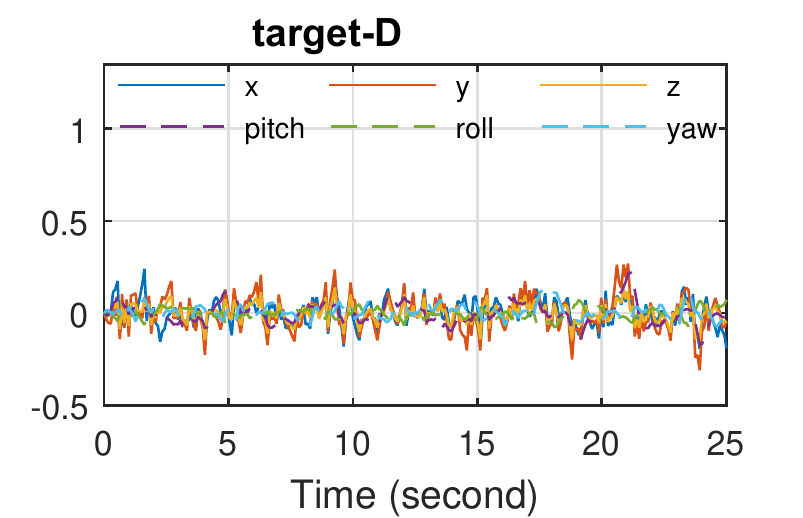}
	}
	\caption{
		Different trajectory planning strategies for real-time motion compensation of volunteer head motion showing strong respiratory coupling.
	}
	\label{fig_vol_respiration}
\end{figure*}

Uncorrected 6DoF head motion of six volunteers was record over a 15 minute tracking period by use of a stereoscopic IR marker tracking system with a 12 Hz sampling rate \cite{belcher2017patient}. In all cases no immobilization was applied to the volunteers. The motion was inputted into the robotic SRS simulation, and platform-D control, target-S planning, and target-D planning were applied, and performances compared. Figure~\ref{fig_vol_respiration} is one example showing a volunteer displaying involuntary drifting and rapid head motion changes due to respiratory coupling. In his case platform-D control fails in meeting the 0.5 mm / 0.2 deg tolerance objective, whereas, the target-D planning is well within tolerance.

\begin{figure*}[ht!]
	\centerline{\sf\footnotesize \underline{\hspace{5mm} uncorrected target motion \hspace{5mm}}}
	\centerline{
		\includegraphics[width=70mm]{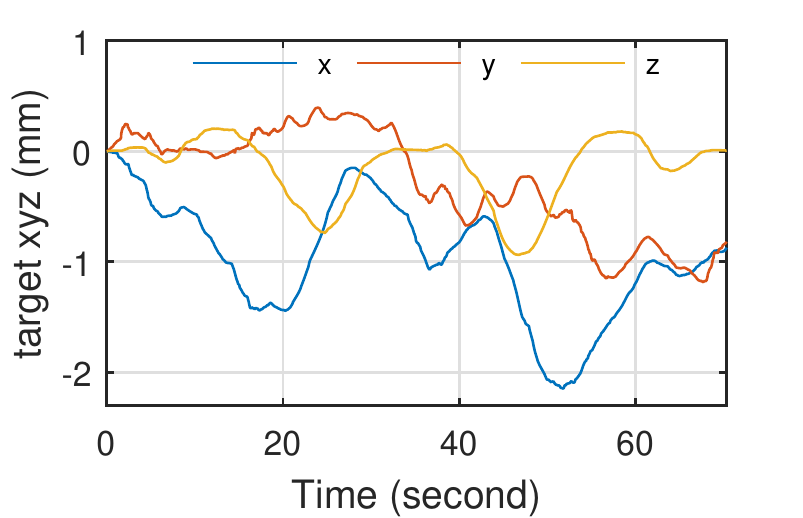}
		\includegraphics[width=70mm]{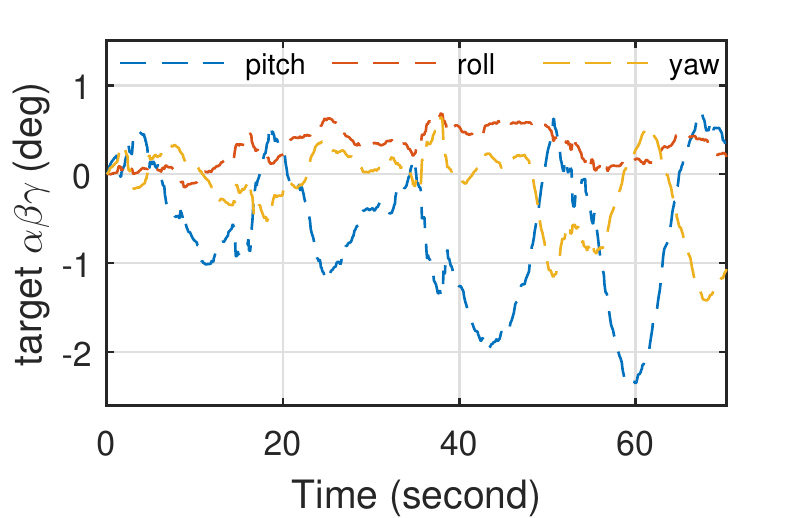}
	}
	\centerline{\sf\footnotesize \underline{\hspace{15mm} corrected target motion by different planning (mm/deg)\hspace{15mm}}}
	\centerline{
		\includegraphics[width=57mm]{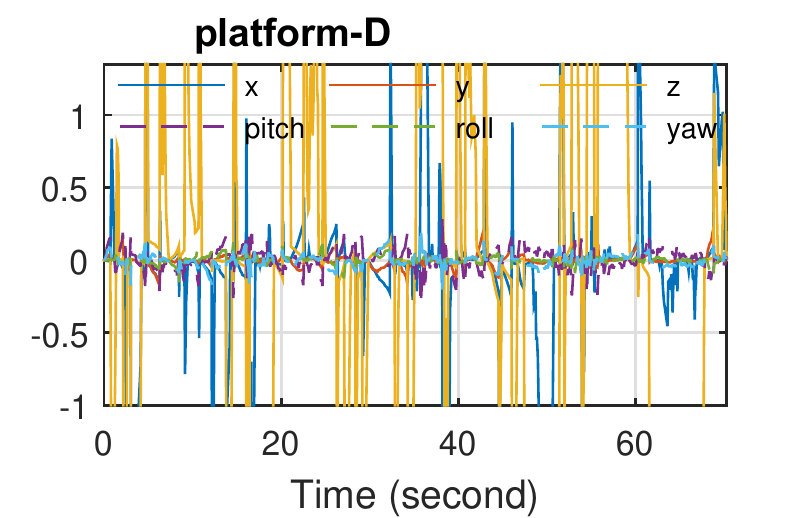}\hspace{-5mm}
		\includegraphics[width=57mm]{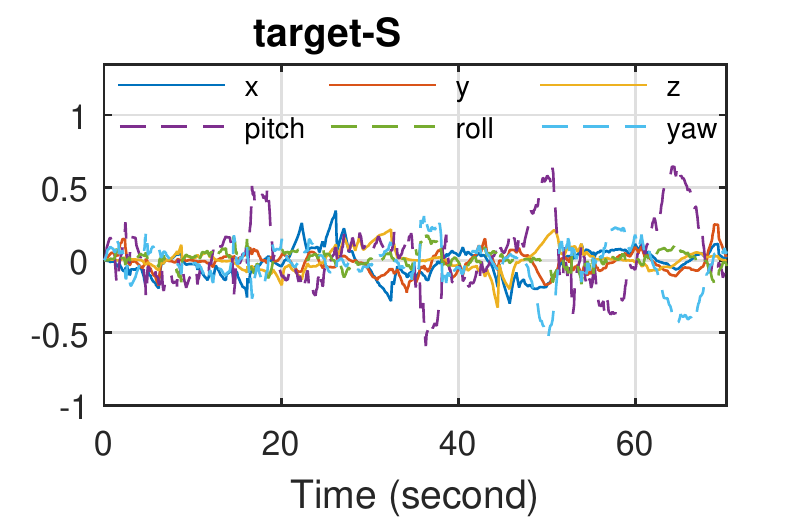}\hspace{-5mm}
		\includegraphics[width=57mm]{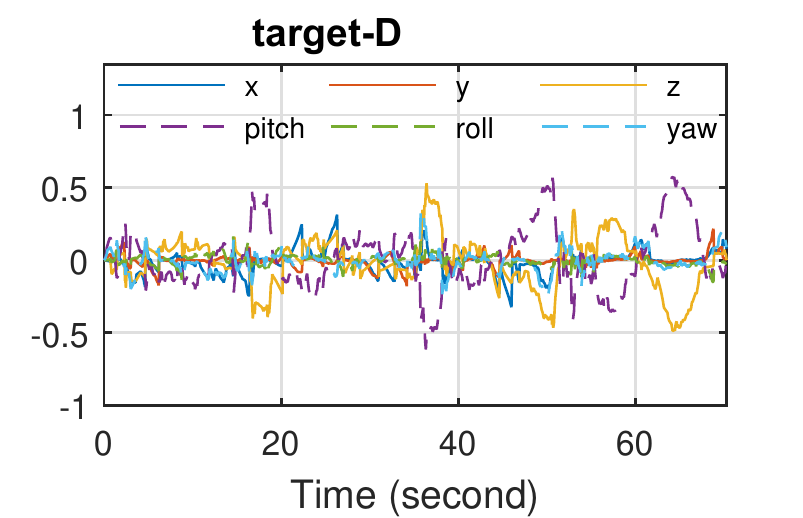}
	}
	\caption{
		Different trajectory control/planning strategies for real-time motion compensation of a prostate tumor using a 6DoF robotic treatment table.
	}
	\label{fig_prostate}
\end{figure*}

To demonstrate the algorithm's applicability to many different types of 6DoF robot systems, Figure~\ref{fig_prostate} shows the results of performing real-time 6D prostate motion compensation using a treatment table. If one were to use a platform-D control to move the platform to the required position as fast as possible, it leads to large intermediate target position errors. On the other hand, both  target-S planning and target-D planning were well within tolerance 0.5 mm / 0.5 deg.

\section*{Discussion}
As the radiation beam is always on during the motion compensation process, it is mandatory that the 6D correction trajectory is optimal both spatially and temporally in order to maximize radiation to the target and minimize unintentional irradiation of healthy tissues. As can be seen in Figure~\ref{fig_sim_DSD1} (b) for a sudden step-like deviation, simply moving the patient support device to the desired position without regard to the target trajectory can result in highly curved trajectories. This is how most modern motion controllers operate, and is an example of an under-constrained system. Although such trajectories are completely acceptable in the vast majority of robot applications, where one is only concerned with moving an object from point A to point B, this is not acceptable in real-time motion compensation using RT as such trajectories may temporally bring OARs, or other sensitive structures, directly into the path of the radiation beam. On the other hand, forcing the target to move along an ideal straight-line, or any well-defined 6D path in target space, can lead to prolonged correction times (target-S plan) as the system becomes over-constrained, in that the robot must move its joints in a fixed way. The use of optimization to explore all potential 6D trajectories by taking into account the robot's joint configuration allows for the various degrees of freedom of the robot to be fully exploited. For example, target-D planning was found to provide a good balance between both spatial and temporal efficiency. As shown in Figure~\ref{fig_sim_DSD1} (c), the 6D target error is quickly reduced, such that in 0.3 s it is approximately 1 mm/deg away from isocenter, compared to $\approx$ 2s using the other approaches.

Typically, in lung tumor motion compensation, or for other targets coupled to respiratory motion, the high velocities necessitate the use of prediction algorithms in order to account for the inherent lag time between actual target position readout and the motion controller response. As shown in Figure~\ref{fig_sim_sin1}, a less optimal correction trajectory will have larger lag times and consequently require more motion prediction. Numerous studies have shown, that the accuracy of prediction algorithms becomes substantially more difficult the further the algorithm must predict into the future \cite{riaz2009predicting}. In this respect, optimization based trajectory planning can improve the accuracy of such motion compensation systems in that an optimal 6D correction path would lower lag time and consequently would require less prediction. Figure~\ref{fig_sim_sin1} compares several different trajectory planning strategies in handling 6D oscillatory motion. As shown in Figure~\ref{fig_sim_sin1} (b), all three different planning strategies reach equilibriums that orbit around the setpoint. However, target-D planning method comes closest to the desired setpoint, which is primary due to its correction path being more optimal than the other methods.

The optimization based motion planning method presented in this work is general in nature and can be applied to any robotic system. The main difference between the compact hexapod and the 6DoF patient treatment table system investigated in this work is the joint space. By defining the table joint space, the method can be fully utilized in order to find optimal 6D trajectories. This is shown in Figure~\ref{fig_prostate}, where real-time motion compensation of prostate motion is performed using the treatment table. However, it should be noted that patient tables are typically designed for general purpose treatment in all parts of the body, and therefore typically have mechanical joint positions inferior of the patient's pelvis in order to prevent significant attenuation or scattering of the radiation beam. When treating targets in the brain, these joints are located far from isocenter and using such a system for SRS motion compensation will pose high demands on the control system in terms of motor synchronization due to mechanical translational and rotational shift. For example, the treatment table investigated in this work (TruePitch, Varian Medical Systems, CA) \cite{schmidhalter2013evaluation}, has a pitch/roll pivot point located approximately 1500 mm away from the patient's head. Therefore, a small 0.05deg pitch error can cause a 1.31 mm displacement error in the vertical direction of the target. On the other hand, site specific robots such as the robotic SRS system are located near the patient's head, has a joint-to-isocenter distance that is approximately 1/4 -- 1/3 that of a table, allowing for lower tolerances on the mechanical systems.

\section*{Conclusion}

A general 6D target trajectory optimization framework for robotic patient motion compensation systems was investigated. Motion planning was formulated as an optimization problem and solved in real-time using a well-known convex optimization algorithm. The method was found to be flexible as it meets various performance requirements such as mechanical robot limits, patient velocities, or other aspects of the system that must operate within fixed limits during the motion compensation process. Both straight-line path and steepest decent trajectories of target error were investigated. In all cases, the steepest decent of the target trajectory was optimal both spatially and temporally.

\section*{Acknowledgments}

This work was funded in part by American Cancer Society grant RSG-13-313-01-CCE.


%
%
%
%
%
%
%

\bibliography{references}

\begin{thebibliography}{10}

\bibitem{deodato2014stereotactic}
Deodato F, Cilla S, Macchia G, Torre G, Caravatta L, Mariano G, et~al.
\newblock Stereotactic radiosurgery (SRS) with volumetric modulated arc therapy
  (VMAT): interim results of a multi-arm phase I trial (DESTROY-2).
\newblock Clinical Oncology. 2014;26(12):748--756.

\bibitem{otto2008volumetric}
Otto K.
\newblock Volumetric modulated arc therapy: IMRT in a single gantry arc.
\newblock Medical physics. 2008;35(1):310--317.

\bibitem{langen2001organ}
Langen K, Jones D.
\newblock Organ motion and its management.
\newblock International Journal of Radiation Oncology* Biology* Physics.
  2001;50(1):265--278.

\bibitem{guckenberger2011potential}
Guckenberger M, Wilbert J, Richter A, Baier K, Flentje M.
\newblock Potential of adaptive radiotherapy to escalate the radiation dose in
  combined radiochemotherapy for locally advanced non--small cell lung cancer.
\newblock International Journal of Radiation Oncology* Biology* Physics.
  2011;79(3):901--908.

\bibitem{van2006conventional}
van~de Bunt L, Van~der Heide UA, Ketelaars M, de~Kort GA,
  J{\"u}rgenliemk-Schulz IM.
\newblock Conventional, conformal, and intensity-modulated radiation therapy
  treatment planning of external beam radiotherapy for cervical cancer: The
  impact of tumor regression.
\newblock International Journal of Radiation Oncology* Biology* Physics.
  2006;64(1):189--196.

\bibitem{kuo2006effect}
Kuo YC, Wu TH, Chung TS, Huang KW, Chao KSC, Su WC, et~al.
\newblock Effect of regression of enlarged neck lymph nodes on radiation doses
  received by parotid glands during intensity-modulated radiotherapy for head
  and neck cancer.
\newblock American journal of clinical oncology. 2006;29(6):600--605.

\bibitem{yan2008developing}
Yan D.
\newblock Developing quality assurance processes for image-guided adaptive
  radiation therapy.
\newblock International Journal of Radiation Oncology* Biology* Physics.
  2008;71(1):S28--S32.

\bibitem{men2010gpu}
Men C, Jia X, Jiang SB.
\newblock GPU-based ultra-fast direct aperture optimization for online adaptive
  radiation therapy.
\newblock Physics in medicine and biology. 2010;55(15):4309.

\bibitem{davies1994ultrasound}
Davies S, Hill A, Holmes R, Halliwell M, Jackson P.
\newblock Ultrasound quantitation of respiratory organ motion in the upper
  abdomen.
\newblock The British journal of radiology. 1994;67(803):1096--1102.

\bibitem{ross1990analysis}
Ross CS, Hussey DH, Pennington EC, Stanford W, Doornbos JF.
\newblock Analysis of movement of intrathoracic neoplasms using ultrafast
  computerized tomography.
\newblock International Journal of Radiation Oncology* Biology* Physics.
  1990;18(3):671--677.

\bibitem{kubo1996respiration}
Kubo HD, Hill BC.
\newblock Respiration gated radiotherapy treatment: a technical study.
\newblock Physics in Medicine \& Biology. 1996;41(1):83.

\bibitem{lohr1999noninvasive}
Lohr F, Debus J, Frank C, Herfarth K, Pastyr O, Rhein B, et~al.
\newblock Noninvasive patient fixation for extracranial stereotactic
  radiotherapy.
\newblock International Journal of Radiation Oncology* Biology* Physics.
  1999;45(2):521--527.

\bibitem{kim2001held}
Kim DJ, Murray BR, Halperin R, Roa WH.
\newblock Held-breath self-gating technique for radiotherapy of non--small-cell
  lung cancer: A feasibility study.
\newblock International Journal of Radiation Oncology* Biology* Physics.
  2001;49(1):43--49.

\bibitem{wiersma2016high}
Wiersma RD, McCabe BP, Belcher AH, Jensen PJ, Smith B, Aydogan B.
\newblock High temporal resolution characterization of gating response time.
\newblock Medical physics. 2016;43(6Part1):2802--2806.

\bibitem{d2005real}
D~D'Souza W, Naqvi SA, Cedric XY.
\newblock Real-time intra-fraction-motion tracking using the treatment couch: a
  feasibility study.
\newblock Physics in medicine and biology. 2005;50(17):4021.

\bibitem{murphy2004tracking}
Murphy MJ.
\newblock Tracking moving organs in real time.
\newblock In: Seminars in radiation oncology. vol.~14. Elsevier; 2004. p.
  91--100.

\bibitem{redpath2008contribution}
Redpath AT, Wright P, Muren LP.
\newblock The contribution of on-line correction for rotational organ motion in
  image-guided radiotherapy of the bladder and prostate.
\newblock Acta Oncologica. 2008;47(7):1367--1372.

\bibitem{wang2008dosimetric}
Wang H, Shiu A, Wang C, O'Daniel J, Mahajan A, Woo S, et~al.
\newblock Dosimetric effect of translational and rotational errors for patients
  undergoing image-guided stereotactic body radiotherapy for spinal metastases.
\newblock International Journal of Radiation Oncology* Biology* Physics.
  2008;71(4):1261--1271.

\bibitem{winey2014geometric}
Winey B, Bussiere M.
\newblock Geometric and dosimetric uncertainties in intracranial stereotatctic
  treatments for multiple nonisocentric lesions.
\newblock Journal of applied clinical medical physics. 2014;15(3):122--132.

\bibitem{roper2015single}
Roper J, Chanyavanich V, Betzel G, Switchenko J, Dhabaan A.
\newblock Single-isocenter multiple-target stereotactic radiosurgery: Risk of
  compromised coverage.
\newblock International Journal of Radiation Oncology* Biology* Physics.
  2015;93(3):540--546.

\bibitem{mancosu2015pitch}
Mancosu P, Reggiori G, Gaudino A, Lobefalo F, Paganini L, Palumbo V, et~al.
\newblock Are pitch and roll compensations required in all pathologies? A data
  analysis of 2945 fractions.
\newblock The British journal of radiology. 2015;88(1055):20150468.

\bibitem{belcher2017patient}
Belcher A.
\newblock Patient motion management with 6dof robotics for frameless and
  maskless stereotatic radiosurgery.
\newblock Ph D Thesis. 2017;.

\bibitem{wilbert2008tumor}
Wilbert J, Meyer J, Baier K, Guckenberger M, Herrmann C, He{\ss} R, et~al.
\newblock Tumor tracking and motion compensation with an adaptive tumor
  tracking system (ATTS): system description and prototype testing.
\newblock Medical physics. 2008;35(9):3911--3921.

\bibitem{menten2012comparison}
Menten MJ, Guckenberger M, Herrmann C, Krau{\ss} A, Nill S, Oelfke U, et~al.
\newblock Comparison of a multileaf collimator tracking system and a robotic
  treatment couch tracking system for organ motion compensation during
  radiotherapy.
\newblock Medical physics. 2012;39(11):7032--7041.

\bibitem{lang2014development}
Lang S, Zeimetz J, Ochsner G, Schmid~Daners M, Riesterer O, Kl{\"o}ck S.
\newblock Development and evaluation of a prototype tracking system using the
  treatment couch.
\newblock Medical physics. 2014;41(2).

\bibitem{buzurovic2011robotic}
Buzurovic I, Huang K, Yu Y, Podder T.
\newblock A robotic approach to 4D real-time tumor tracking for radiotherapy.
\newblock Physics in medicine and biology. 2011;56(5):1299.

\bibitem{haas2012couch}
Haas OC, Skworcow P, Paluszczyszyn D, Sahih A, Ruta M, Mills JA.
\newblock Couch-based motion compensation: modelling, simulation and real-time
  experiments.
\newblock Physics in medicine and biology. 2012;57(18):5787.

\bibitem{herrmann2011model}
Herrmann C, Ma L, Schilling K.
\newblock Model predictive control for tumor motion compensation in robot
  assisted radiotherapy.
\newblock IFAC Proceedings Volumes. 2011;44(1):5968--5973.

\bibitem{buzurovic2012robust}
Buzurovic I, Debeljkovic DL.
\newblock Robust control for parallel robotic platforms.
\newblock In: Intelligent Engineering Systems (INES), 2012 IEEE 16th
  International Conference on. IEEE; 2012. p. 45--50.

\bibitem{liu2015robotic}
Liu X, Belcher AH, Grelewicz Z, Wiersma RD.
\newblock Robotic real-time translational and rotational head motion correction
  during frameless stereotactic radiosurgery.
\newblock Medical Physics. 2015;42(6):2757--2763.

\bibitem{liu2015roboticACC}
Liu X, Belcher AH, Grelewicz Z, Wiersma RD.
\newblock Robotic stage for head motion correction in stereotactic
  radiosurgery.
\newblock In: 2015 American Control Conference (ACC);.

\bibitem{belcher2017towards}
Belcher AH, Liu X, Chmura S, Yenice K, Wiersma RD.
\newblock Towards frameless maskless SRS through real-time 6DoF robotic motion
  compensation.
\newblock Physics in Medicine \& Biology. 2017;62(23):9054.

\bibitem{wiersma2009development}
Wiersma RD, Wen Z, Sadinski M, Farrey K, Yenice KM.
\newblock Development of a frameless stereotactic radiosurgery system based on
  real-time 6D position monitoring and adaptive head motion compensation.
\newblock Physics in medicine and biology. 2009;55(2):389.

\bibitem{liu2018general}
Liu X, Wiersma R.
\newblock A General Patient Motion Compensation in Robotic Systems Using
  Optimization Based 6DoF Trajectory Planning.
\newblock In: MEDICAL PHYSICS. vol.~45; 2018. p. E644--E644.

\bibitem{belcher2016spatial}
Belcher AH, Liu X, Grelewicz Z, Wiersma RD.
\newblock Spatial and rotational quality assurance of 6DOF patient tracking
  systems.
\newblock Medical Physics. 2016;43(6):2785--2793.

\bibitem{belcher2014development}
Belcher AH, Liu X, Grelewicz Z, Pearson E, Wiersma RD.
\newblock Development of a 6DOF robotic motion phantom for radiation therapy.
\newblock Medical physics. 2014;41(12):121704.

\bibitem{wiersma2013spatial}
Wiersma RD, Tomarken S, Grelewicz Z, Belcher AH, Kang H.
\newblock Spatial and temporal performance of 3D optical surface imaging for
  real-time head position tracking.
\newblock Medical physics. 2013;40(11):111712.

\bibitem{grelewicz2014combined}
Grelewicz Z, Wiersma RD.
\newblock Combined MV+ kV inverse treatment planning for optimal kV dose
  incorporation in IGRT.
\newblock Physics in medicine and biology. 2014;59(7):1607.

\bibitem{siciliano2010robotics}
Siciliano B, Sciavicco L, Villani L, Oriolo G.
\newblock Robotics: modelling, planning and control.
\newblock Springer Science \& Business Media; 2010.

\bibitem{morales2011remark}
Morales JL, Nocedal J.
\newblock Remark on “Algorithm 778: L-BFGS-B: Fortran subroutines for
  large-scale bound constrained optimization”.
\newblock ACM Transactions on Mathematical Software (TOMS). 2011;38(1):7.

\bibitem{parikh2014block}
Parikh N, Boyd S.
\newblock Block splitting for distributed optimization.
\newblock Mathematical Programming Computation. 2014;6(1):77--102.

\bibitem{fougner2015parameter}
Fougner C, Boyd S.
\newblock Parameter selection and pre-conditioning for a graph form solver.
\newblock Online at http://stanfordedu/~boyd/papershtml. 2015;.

\bibitem{liu2016constrained}
Liu X, Belcher AH, Grelewicz Z, Wiersma RD.
\newblock Constrained quadratic optimization for radiation treatment planning
  by use of graph form ADMM.
\newblock In: American Control Conference (ACC), 2016. American Automatic
  Control Council (AACC); 2016. p. 5599--5604.

\bibitem{liu2017use}
Liu X, Pelizzari C, Belcher AH, Grelewicz Z, Wiersma RD.
\newblock Use of proximal operator graph solver for radiation therapy inverse
  treatment planning.
\newblock Medical Physics. 2017;44(4):1246--1256.

\bibitem{tehrani2013real}
Tehrani JN, T~O’Brien R, Poulsen PR, Keall P.
\newblock Real-time estimation of prostate tumor rotation and translation with
  a kV imaging system based on an iterative closest point algorithm.
\newblock Physics in Medicine \& Biology. 2013;58(23):8517.

\bibitem{riaz2009predicting}
Riaz N, Shanker P, Wiersma R, Gudmundsson O, Mao W, Widrow B, et~al.
\newblock Predicting respiratory tumor motion with multi-dimensional adaptive
  filters and support vector regression.
\newblock Physics in Medicine \& Biology. 2009;54(19):5735.

\bibitem{schmidhalter2013evaluation}
Schmidhalter D, Fix M, Wyss M, Schaer N, Munro P, Scheib S, et~al.
\newblock Evaluation of a new six degrees of freedom couch for radiation
  therapy.
\newblock Medical physics. 2013;40(11).

\end{thebibliography}


\end{document}